\documentclass[oneside,a4paper,english,links,12pt]{article}

\pdfoutput=1
\usepackage{amsmath}
\usepackage{authblk}
\usepackage{natbib}

\usepackage{amssymb}
\usepackage{subfigure}
\usepackage{graphicx}
\usepackage{amsfonts}

\usepackage{calc}
\usepackage{ifpdf}
\usepackage{indentfirst}
\usepackage{authblk}
\usepackage{babel}
\usepackage{color}
\usepackage{hyperref}
\usepackage{nameref}
\usepackage{url}
\usepackage{times}
\usepackage{fontenc}
\usepackage{hyperref}

\usepackage{vmargin}

\setpapersize{A4}
\setmargins{2.5cm}  
{1.5cm}             
{16.5cm}            
{23.42cm}           
{12pt}              
{1cm}               
{0pt}               
{2cm}               

\def\brokenline{{$- - - $}}

\def\diamonds{\normalsize{$\diamond\cdot\diamond\cdot\diamond$}\small}
\def\circs{\normalsize{$\circ \cdot \circ \cdot \circ\;$\small}}

\def\boxs{\scriptsize{$\Box \cdot \Box \cdot \Box $ \small}}

\def\pluss{\scriptsize{$+\cdot+\cdot+$}\small} 

\title{Using LSTM Predictions for RANS Simulations}

\author{Hugo D. Pasinato}

\affil{Universidad Tec. Nacional, FRP, Argentina}

\affil{hugopasinato@frp.utn.edu.ar; hugodariop@hotmail.com}

\date{}

\begin{document}

\maketitle 

\noindent This study constitutes the second phase of a research
endeavor aimed at evaluating the feasibility of employing Long
Short-Term Memory (LSTM) neural networks as a replacement for
Reynolds-Averaged Navier-Stokes (RANS) turbulence models.

In the initial phase of this investigation (titled {\it Modeling
Turbulent Flows with LSTM Neural Networks}, arXiv:2307.13784v1
[physics.flu-dyn] 25 Jul 2023), the application of an LSTM-based
recurrent neural network (RNN) as an alternative to traditional RANS
models was demonstrated. LSTM models were used to predict shear
Reynolds stresses in both developed and developing turbulent channel
flows, and these predictions were propagated through RANS simulations
to obtain mean flow fields of turbulent flows. A comparative analysis
was conducted, juxtaposing the LSTM results from computational fluid
dynamics (CFD) simulations with outcomes from the $\kappa-\epsilon$
model and data from direct numerical simulations (DNS). These initial
findings indicated promising performance of the LSTM approach.

This second phase delves further into the challenges encountered and
presents robust solutions. Additionally, new results are provided,
demonstrating the efficacy of the LSTM model in predicting turbulent
behavior in perturbed flows. While the overall study serves as a
proof-of-concept for the application of LSTM networks in RANS
turbulence modeling, this phase offers compelling evidence of its
potential in handling more complex flow scenarios.


\section{Introduction}

The objective of this research is to investigate the use of Recurrent
Neural Networks (RNNs), specifically Long Short-Term Memory (LSTM)
models, as a substitute for traditional turbulence models within the
framework of Reynolds-averaged Navier-Stokes (RANS) equations
\cite{pope2000}. These equations, expressed in dimensionless form
using tensor notation, represent the conservation of mass:

\begin{equation}\label{eq101}
\frac{\partial U_i}{\partial x_i } =0,
\end{equation}

\noindent and the momentum equations:

\begin{equation}\label{eq102}
\frac{\partial U_i}{\partial t}+ \frac{\partial (U_j U_i)}{\partial
  x_j} = -\frac{\partial P}{\partial x_i} + \frac{1}{Re_{\tau}}
\frac{\partial^2 U_i}{\partial x_j \partial x_j} - \frac{\partial
  \langle u_j u_i \rangle }{\partial x_j},
\end{equation}

\noindent where \( U_i \) (corresponding to \( U \), \( V \), and \( W
\)) represents the mean velocity, \( P \) the mean pressure, and \( x
\) and \( t \) are the spatial and temporal coordinates,
respectively. The friction Reynolds number, \( Re_{\tau} \), is
defined as \( u_{\tau} \delta / \nu \), where all variables are
non-dimensionalized using the friction velocity \( u_{\tau} \) and
half the distance between the walls \( \delta \), serving as the
characteristic velocity and length scales.

The last term on the right-hand side of equation (\ref{eq102}) is the
Reynolds stress tensor \( \langle u_j u_i \rangle \), which arises
from time-averaging the original Navier-Stokes equations and
introduces new unknowns into the system. Since the Reynolds stress
tensor is symmetric, solving the RANS equations necessitates
addressing six additional unknowns, alongside the velocity components
\( U_i \) and the pressure \( P \).

To resolve the RANS equations, a method for determining these six
unknowns must be incorporated, as they are needed to compute the
velocities \( U \), \( V \), and \( W \), along with the pressure \( P
\), derived from the conservation equations of momentum and
mass. These methods are collectively referred to as RANS turbulence
modeling, with the Kappa-Epsilon model being a well-known
example. Such models typically employ a pseudo viscosity that accounts
for the effects of turbulence, drawing an analogy between the behavior
of a Newtonian fluid and a turbulent flow. By incorporating this
pseudo viscosity, these models aim to replicate the dynamics of
turbulent flows in a manner similar to the behavior of Newtonian
fluids.

The current study aims to replace the turbulence model within the RANS
framework with an LSTM-based recurrent neural network. In a previous
preprint \cite{pasinato2023}, the fundamental concepts of this
approach and initial results were presented. This preprint provides a
detailed description of how the LSTM's predictions are propagated. The
following sections outline the architecture of the LSTM models
employed in the study (§2), followed by a description of the DNS data
used for model training and adjustments (§3). Numerical details of the
RANS simulations are provided in §4, while the results are presented
in §5, with conclusions drawn in §6.

\section{LSTM Models}

One architecture of LSTM RNN is the sequence-by-sequence (seqxseq)
architecture. In this format, a sequence in time of input-output pairs
is trained (\cite{schmidhuber, sutskever}). However, to apply this
architecture within RANS simulations, a fundamental adaptation was
necessary. Most turbulent flows analyzed in RANS simulations are
statistically stationary, meaning the Reynolds-averaged values along
the x-direction at a constant distance from the boundary form a
spatial sequence—from upstream to downstream—. Therefore, in this
study, the LSTM neural network was trained using spatial
sequence-by-sequence data instead of time-based sequences, as
explained in the first part of the study (\cite{pasinato2023}). Each
sequence comprised 32 consecutive input-output pairs corresponding to
nodes in the RANS simulations along the longitudinal direction,
spanning the entire physical domain. The data were normalized between
0 and 1, using the global maximum and minimum values from the dataset.

Once the Reynolds shear stresses are predicted, new challenges arise
regarding their integration into the RANS simulations. One of the key
difficulties in employing neural networks to predict Reynolds stresses
in RANS simulations lies in how these turbulent stress values are
incorporated into computational fluid dynamics (CFD) codes to produce
a robust solution for turbulent flow. While neural networks trained on
Direct Numerical Simulation (DNS) databases can accurately predict
Reynolds stresses, improvements in mean flow fields are not
guaranteed, as highlighted by \cite{wu2019}. Wu and colleagues
explored potential ill-conditioning of the RANS equations when
explicit Reynolds stresses are included in mean velocity
calculations. Even when DNS-derived Reynolds stresses are used in RANS
simulations at high Reynolds numbers, significant errors in velocity
predictions can still occur \cite{thompson2016}.

Although the present study focuses on turbulent flows at moderately
low Reynolds numbers, addressing the explicit incorporation of
stresses into the RANS equations required consideration of two key
aspects to develop robust predictions. First, it was acknowledged that
the fragility of neural networks often correlates with their capacity,
defined by the number of parameters. Thus, the initial step involved
reducing the number of parameters without risking
underfitting—essentially, designing a parsimonious LSTM
model. Striking the right balance between complexity and simplicity
was essential, ensuring that only the necessary parameters were
included to adequately represent the phenomena, thereby avoiding
unnecessary redundancy or overfitting.

The second aspect involved the use of bidirectional wrapping
architectures. As previously noted, turbulence is not a local
phenomenon; the transport of momentum by turbulence at a given point
in space depends on the flow characteristics both upstream and
downstream of this point. Therefore, based on prior experience, a
strategy to improve the stability of using explicit Reynolds stresses
in the RANS equations involved employing LSTM models with
bidirectional wrapping architectures.

\begin{table}[ht]
\begin{center}
\begin{tabular}{lllll}
\hline \hline
Name            & Stacked LSTM & Parameters & Features     & Prediction\\
\hline
1XL110DL        &   0   & 491   & $S_{12}^+$                                   & $\langle uv \rangle^+ $\\
2XL120DL        &   0   &1861   & $S_{12}^+; Y$                                & $\langle uv \rangle^+ $\\
2XBL13BL23BL33DL&   3   & 631   & $S_{12}^+; Y$                                & $\langle uv \rangle^+ $\\
3XBL13BL23BL33DL&   3   & 655   & $S_{12}^+; Y; Re_{\tau}$                     & $\langle uv \rangle^+ $\\
4XBL13BL23BL33DL&   3   & 679   & $Re_{\tau}; S_{11}^+; S_{12}^+; Y$           & $\langle uv \rangle^+ $\\
4XBL14BL24DL    &   3   & 713   & $Re_{\tau}; S_{11}^+; S_{12}^+; Y$           & $\langle uv \rangle^+ $\\
5XBL13BL23BL33DL&   3   & 703   & $Re_{\tau}; S_{11}^+; S_{12}^+; S_{22}^+; Y$ & $\langle uv \rangle^+ $\\
\hline
\end{tabular}
\end{center}
\caption[table2]{\label{table2}\small LSTM models. }
\end{table}

Table \ref{table2} presents some of the LSTM architectures employed in
the second phase of this study. For example, $4XBL13BL23BL33DL$
denotes an LSTM model with an input vector comprising four features
and three stacked LSTM layers, each with three memory units and
bidirectional wrapping, and a dense layer at the output. The features
$S_{11}^+$, etc., represent components of the dimensionless mean
rate-of-strain tensor $S_{ij}$, calculated as \( S_{ij}^+ = ({1}/{2})
\left({\partial U^+_i}/{\partial x^+_j} + {\partial U^+_j}/{\partial
  x^+_i} \right) \). Additionally, $Y$ and \( Re_{\tau} = {u_{\tau}
  \delta \rho}/{\mu} \) denote the dimensionless wall distance
($y/\delta$) and the friction Reynolds number, respectively. It is
important to note that the axes in $S_{ij}$ are non-dimensionalized
differently from the wall distance $Y$.

The LSTM model, like other deep learning models, involves three
primary hyperparameters: (a) the number of stacked layers (or stacked
LSTM cells), (b) the number of memory units in each LSTM layer, and
(c) the learning rate. These hyperparameters were selected in this
study through trial and error.

As discussed in \cite{pasinato2023}, another critical aspect of a
neural network model is its universality. The appropriate balance
between universality and simplicity must be addressed—how specialized
and simple the model can be while maintaining performance. Developing
a fully universal NN model would require a large number of adjustable
parameters to capture all possible flow characteristics. However,
fine-tuning such a model poses significant challenges, and integrating
a neural network with a high parameter count into CFD software demands
substantial computational resources. Therefore, this study sought to
employ an RNN with a reduced parameter count to facilitate rapid
predictions within a RANS simulation.

At this stage, the computational time of CFD simulations using an LSTM
network was not the primary focus. The main objective was to develop
an effective tool to replace traditional RANS turbulence models with a
neural network. For comparison purposes, the computational time of a
standard CFD simulation using the $\kappa-\epsilon$ model
\cite{pope2000} was used as a reference. The goal was to employ LSTM
architectures with no more than approximately 700 parameters, as LSTM
models with this parameter count take approximately the same
computational time as the $\kappa-\epsilon$ model. Although the LSTM
model allows for larger time steps, each time step of the
$\kappa-\epsilon$ model remains faster than those of the LSTM model.

In the second phase of this research, all software required for tuning
the LSTM parameters was developed in Python using the Keras-TensorFlow
libraries \cite{tensorflow2015-whitepaper, brownlee}. The LSTMforRANS
repository, available on GitHub at
(https://github.com/hugodariopasinato/LSTM-for-RANS), contains
essential resources including the LSTM neural network model
\textit{4XBL13BL23BL33DL}, the maxima and minima from the DNS dataset,
and FORTRAN subroutines designed to load the LSTM model and use it for
predictions. These subroutines can be integrated into CFD codes to
enable RANS simulations incorporating LSTM-based predictions of
Reynolds stress.

\section{Training, testing, and validation data}

\begin{table}[ht]
\begin{center}
\begin{tabular}{lllccc}
\hline \hline
Name     &$Re_{\tau}$ & Perturbation & Parameter                    & Purpose     \\
\hline
Developed& 150; 300; 600&        &           & training   \\
Developed& 200      &        &                       & validation  \\
B1506    & 150      & Blowing &$v^+=0.6$                         & training    \\
S1506    & 150      & Suction &$v^+=0.6$                         & training    \\
APGS30025  & 300      & APGS    &$\partial P^+/\partial x^+=+0.25$ & training    \\
FPGS30025  & 300      & FPGS    &$\partial P^+/\partial x^+=-0.25$ & training    \\
WF150    & 150      & Frictionless wall &                        & validation   \\
WF300    & 300      & Frictionless wall &                        & validation  \\
WFB3003  & 300      & Frictionless wall/Blowing &$v^+=0.3$       & validation \\
WFAPGS15020& 150      & Frictionless wall/APGS &$\partial P^+/\partial x^+=+0.20$ & validation    \\
WFAPGS20025& 200      & Frictionless wall/APGS &$\partial P^+/\partial x^+=+0.25$ & validation    \\
WFFPGS20025& 200      & Frictionless wall/APGS &$\partial P^+/\partial x^+=-0.25$ & validation    \\
WFFPGS30010& 300      & Frictionless wall/APGS &$\partial P^+/\partial x^+=-0.10$ & validation    \\
WFAPGS30010& 300      & Frictionless wall/APGS &$\partial P^+/\partial x^+=-0.10$ & validation    \\
WFS3003   & 300      & Frictionless wall/Suction &$v^+=0.3$        & validation \\
BS1503   & 150      & Blowing/Suction &$v^+=0.3$                    & validation\\
SB3003   & 300      & Suction/Blowing &$v^+=0.3$                   & validation\\
\hline
\end{tabular}
\end{center}
\caption[table1]{\label{table1}\small Dataset used for training,
  testing, or validation purposes. $v^+$ represents the dimensionless
  wall-normal velocity at the wall for blowing or suction; $\partial
  P^+/\partial x^+$ denotes the pressure gradient step. "APGS" and
  "FPGS" refer to adverse and favorable pressure gradient steps,
  respectively. "WFAPGS" and "WFFPGS" refer to frictionless walls
  combined with APGS or FPGS. "BS" represents blowing followed by
  suction, and "SB" represents the reverse.}
\end{table}

The DNS database of turbulent flows used for training, testing, and
validation in this study was generated using a DNS numerical code that
solves the incompressible momentum equations. These equations were
discretized using a second-order accurate central-difference
scheme. The Poisson equation for the pressure field was
Fourier-transformed along the streamwise and spanwise directions, and
the resulting tridiagonal systems were solved directly at each time
step. The flow field evolution was computed using a fractional-step
method, with the Crank-Nicolson second-order scheme applied to viscous
terms, and the Adams-Bashforth scheme to non-linear terms. Further
numerical details on the DNS calculations are available in
\cite{pasinato2012} and \cite{pasinato2014}.

The dataset consists of statistically stationary 2D turbulent channel
flows at moderately low Reynolds numbers, which serves as a
proof-of-concept for the application of LSTM networks in RANS
turbulence modeling. The dataset includes developed turbulent channel
flows for friction Reynolds numbers, $R_{\tau}$, of 150, 200, 300, and
600. The minimum friction Reynolds number in this study, $Re_{\tau} =
150$, corresponds to $Re_{\delta} \simeq 2900$, where $Re_{\delta} = U
2 \delta/\nu \simeq (Re_{\tau}/0.1406)^{(8/7)}$, with $U$ being the
mean velocity (where the $Re_{\delta, crit} \simeq
1400$). Additionally, the dataset includes perturbed channel flows
subjected to wall blowing or suction, adverse (APGS) or favorable
pressure gradient steps (FPGS), changes in wall friction (WF), and
combinations of these perturbations, for $R_{\tau}$ values of 150,
200, 300, and 600. Table \ref{table1} provides a detailed list of
these flows, including perturbation parameters and their respective
purposes.

For instance, the case "WFB3003" represents a turbulent flow initially
developed with $R_{\tau}=300$ that is perturbed in a narrow region of
width $W^+=206$ by wall blowing combined with a frictionless wall. In
contrast, the case "BS2003" refers to a flow initially developed with
$R_{\tau}=200$, which is perturbed in two contiguous regions of width
$W^+=206$ by blowing followed by suction, with $v^+=0.3$, resulting in
a total perturbation region of $W^+=412$.

To understand the turbulence transport mechanisms in these perturbed
cases in a RANS simulation, it is important to examine the impact on
longitudinal momentum transport (x-direction). The transport of
longitudinal momentum between the central region and the wall, which
in a perturbed flow occurs via mean convection, turbulent flux, and
molecular viscosity, is altered from the developed flow conditions. In
developed flows, momentum transport towards the wall is driven
primarily by the Reynolds stress, $\langle uv \rangle$, with no mean
convection towards or away from the wall, and molecular viscosity
becoming significant only in the near-wall region.

Each type of perturbation modifies the transport of longitudinal
momentum between the central region and the wall. Perturbations such
as blowing, suction, pressure gradient steps, and changes in wall
friction modify $\langle uv \rangle$ and introduce mean convection
towards or away from the wall. When a perturbation disrupts natural
turbulent momentum transport, turbulence either intensifies or
weakens, depending on whether the perturbation aids or opposes the
momentum transfer.

In cases where perturbations introduce mean wall-normal convection
that opposes turbulent fluxes (e.g., blowing, adverse pressure
gradients, or increased wall friction), turbulence intensifies to
compensate for the momentum deficit near the wall. Conversely, when
perturbations assist momentum transport towards the wall (e.g.,
suction, favorable pressure gradients, or reduced wall friction),
turbulent transport decreases. A well-known example is the
relaminarization that occurs when a strong favorable pressure gradient
is applied to a previously developed turbulent flow.

For validation purposes, the types and parameters of perturbations
were adjusted. In the training flows, pressure gradient steps were
introduced in the buffer region, whereas in the validation cases, they
were applied in the logarithmic region. In both scenarios, the
perturbations were confined to a narrow longitudinal region of width
$W^+=206$. The combination of multiple perturbations can also
significantly alter turbulent behavior. For example, a favorable
pressure gradient step, which increases skin friction by drawing fluid
towards the wall, behaves differently when applied alongside a
frictionless wall. Similarly, combining blowing and suction leads to
distinct effects compared to applying them individually. A suction
region followed by a blowing region, for instance, induces sharp
curvature in the mean flow streamlines, producing patterns that differ
from those caused by either perturbation alone.

Future studies will require the development of new turbulent flows
under varied boundary conditions and at higher Reynolds numbers to
achieve more comprehensive validation. However, to ensure practical
applicability in CFD codes, it is crucial to focus on developing
neural networks tailored to specific types of turbulence, rather than
pursuing universal models.

As described in Table \ref{table1}, the "training data" consists of
DNS data used to adjust the LSTM model parameters. This data is
further split into "training" (80\%) and "testing" (20\%) subsets
during the training process. The "validation data" refers to new DNS
data used exclusively to compare DNS results with RANS simulations
where LSTM predictions were incorporated.

\section{Numerical Details of RANS Simulations}

Each RANS simulation of perturbed turbulent flow (mirroring the
approach used in the DNS simulations of perturbed flow) was conducted
with two parallel simulations. Both simulations represent channel flow
with the same $Re_{\tau}$, physical domain, and grid. The first
simulation, using periodic boundary conditions, provided the inlet
boundary conditions for the second simulation. This second simulation,
which includes the perturbed (developing) flow, used the developed
turbulent flow from the first simulation as its inlet condition, while
convection boundary conditions were applied at the outlet. The width
of the perturbation region, or slot, was fixed at $W^+ = 206$ for all
cases, and the distance from the inlet to the perturbation region was
approximately $X^+ \simeq 600$. In cases where blowing was combined
with suction (or vice versa), the total perturbation width was
doubled, reaching $2W^+ = 412$.

All RANS simulations, both for the $\kappa-\epsilon$ model and for
LSTM propagation, were conducted using the same computational domain
and grid, with the only difference being the time step. The
$\kappa-\epsilon$ model required a time step four times smaller than
that used for the LSTM's prediction propagation, due to the
constraints imposed by solving the $\kappa$ and $\epsilon$
conservation equations numerically.

Both LSTM propagation and the $\kappa-\epsilon$ simulations were
carried out within a physical domain scaled according to the
$Re_{\tau}$ number. For example, at $Re_{\tau} = 150$, the
computational domain was $4\pi \delta \times 2\delta \times
\frac{4}{3}\pi \delta$. For higher Reynolds numbers, this domain was
adjusted accordingly. All RANS results presented here used a numerical
grid of $32 \times 32 \times 32$, although a finer grid of $64 \times
64 \times 64$ was also tested. To employ a $64 \times 64 \times 64$
grid, an LSTM with sequences of 64 elements would be required; using
an LSTM with sequences of only 32 elements on a grid of 64 nodes can
lead to inconsistencies at grid junctions. For $Re_{\tau} = 150$, the
grid resolution was $\Delta x^+ = 88$, $\Delta z^+ = 39.2$, $\Delta
y^+_{max} = 52.8$, and $\Delta y^+_{min} = 3$. A non-uniform mesh,
distributed using a hyperbolic tangent function in the wall-normal
direction, ensured that the $y^+$ value at the first cell center
remained below 3. Near the wall, a van Driest damping function was
applied with the $\kappa-\epsilon$ model. Given the low Reynolds
number flows considered in this study, this wall-modeling approach is
deemed appropriate.

\section{Results}

\begin{figure}[ht] 
{\par\centering
  \resizebox*{0.70\columnwidth}{!}{\includegraphics{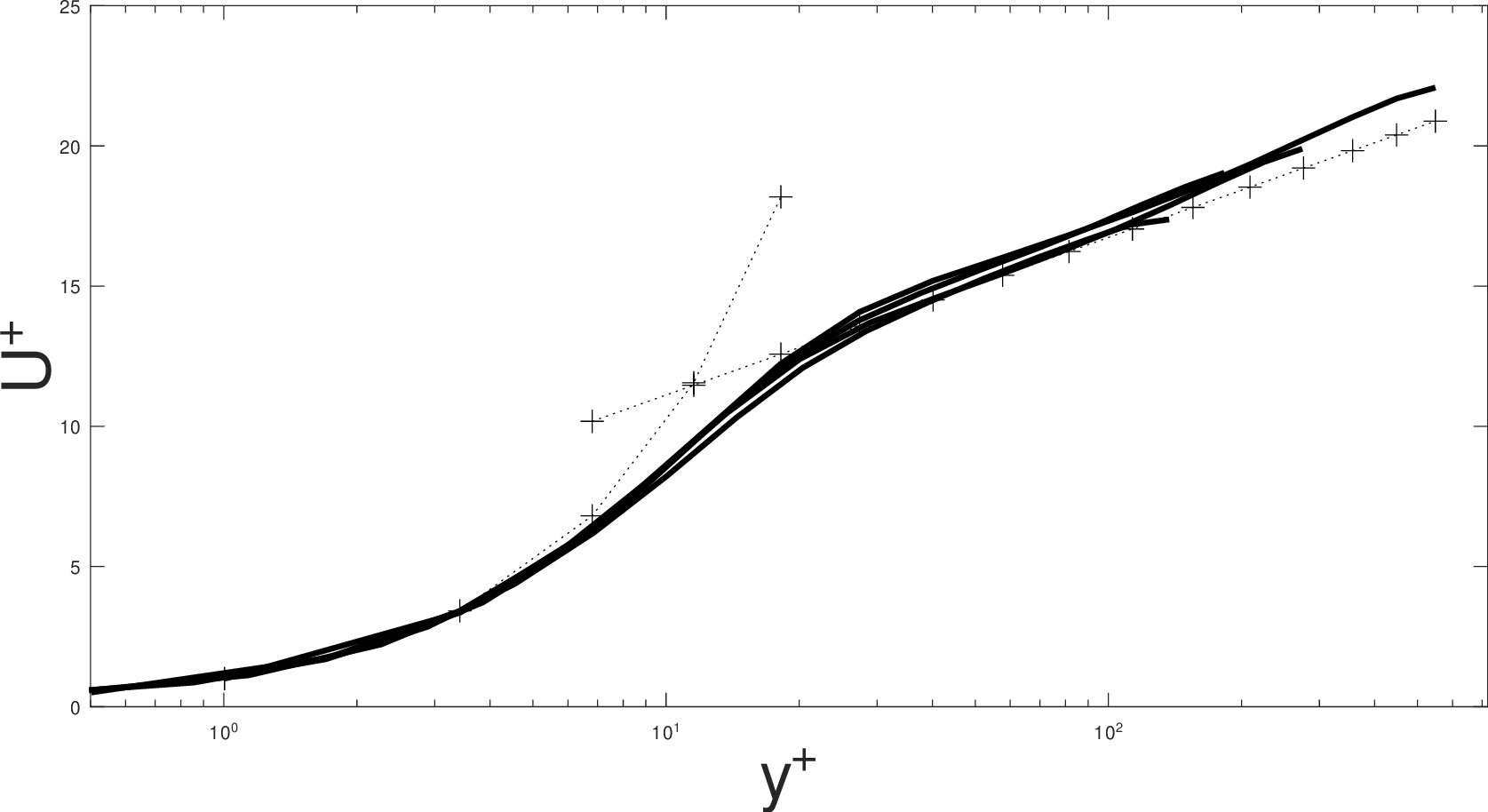}}
  \par}
\caption{\small{Dimensionless longitudinal velocity $U^+$ for various
    Reynolds numbers ($Re_{\tau}=150$, $200$, $300$, $600$) predicted
    by the LSTM model for developed channel flows. Solid line: LSTM;
    \pluss, \normalsize Law of the Wall, $U^+ = ({1}/{0.41})\ln(y^+) +
    5.5$.}}
\label{figure1} 
\end{figure}

\begin{figure}[ht] 
{\par\centering
  \resizebox*{0.70\columnwidth}{!}{\includegraphics{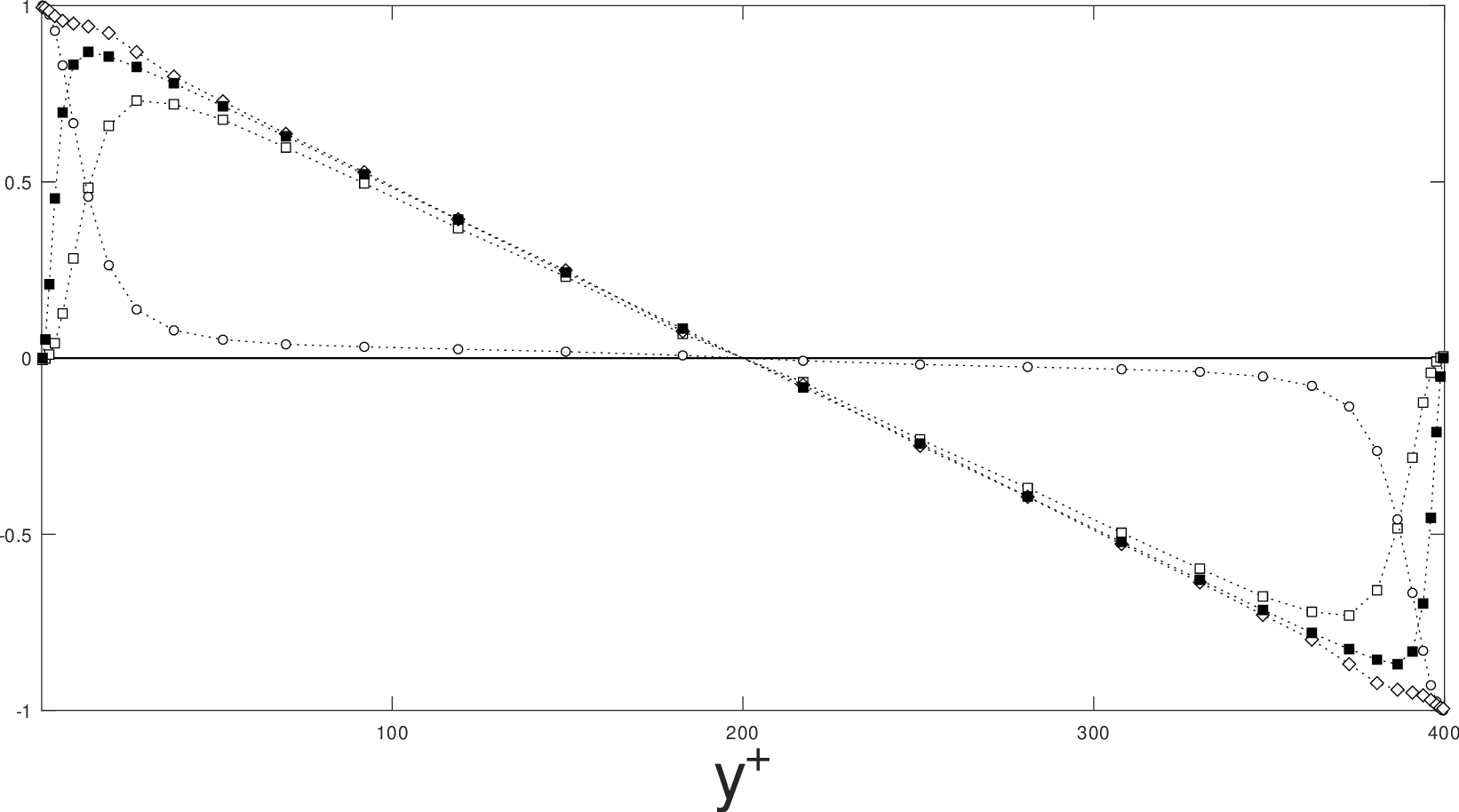}}
  \par}
\caption{\small{Shear Reynolds stresses predicted by the LSTM model
    for a developed channel flow with $Re_{\tau}=200$. \diamonds
    \normalsize Total; \circs \normalsize Molecular; \boxs \normalsize
    $\langle uv \rangle^+$; filled \boxs \normalsize $\langle uv
    \rangle^+$ for $Re_{\tau}=600$ (using the same $y^+$ scale as for
    $Re_{\tau}=200$).}}
\label{figure2} 
\end{figure}

The LSTM models used in this study predict only the $\langle uv
\rangle$ component of the Reynolds stress tensor. Therefore, in the
RANS simulations where LSTM predictions are propagated, all other
Reynolds stress components are set to zero. In general 3D turbulent
flows, the Reynolds stresses, which result from time-averaging the
Navier-Stokes equations, form a 9-component tensor. Since this tensor
is symmetric, only six independent components need to be calculated by
a RANS turbulence model. These include three normal stresses: $\langle
uu \rangle$, $\langle vv \rangle$, and $\langle ww \rangle$, which
represent turbulent flux transport in the x, y, and z directions,
respectively. The remaining three components are shear stresses:
$\langle uv \rangle$, $\langle uw \rangle$, and $\langle vw \rangle$,
which are off-diagonal components of the tensor. Specifically,
$\langle uv \rangle$ governs the turbulent transport of x-direction
momentum in the y-direction, $\langle vw \rangle$ governs y-direction
momentum in the z-direction, and $\langle uw \rangle$ governs
z-direction momentum in the x-direction.

\begin{figure}[ht] 
{\par\centering
  \resizebox*{0.70\columnwidth}{!}{\includegraphics{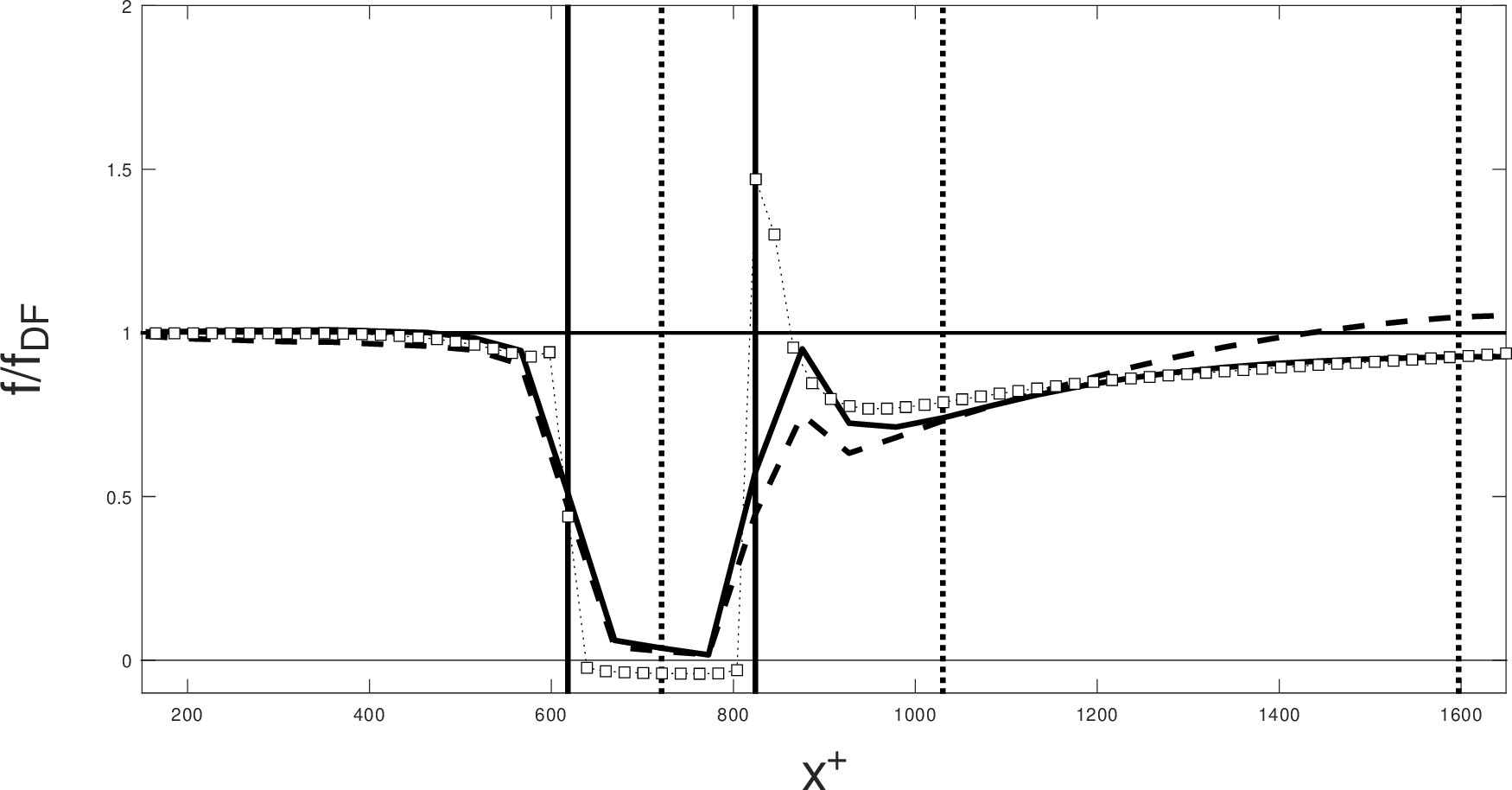}}
  \par}
\caption{\small{Normalized skin-friction coefficient $f/f_{DF}$ for a
    channel flow with $Re_{\tau}=200$, perturbed by a frictionless
    wall and an adverse pressure gradient step in a narrow region of
    width $W^+ = 206$. Solid vertical lines indicate the boundaries of
    the perturbation region, while dashed vertical lines represent its
    center and at distances of $W^+$ and $5W^+$ from its
    end. \brokenline, \normalsize $\kappa-\epsilon$ model; \boxs
    \normalsize, DNS; solid line, LSTM.}}
\label{figure3} 
\end{figure}

\begin{figure}[tbp] 
{\par\centering
  \resizebox*{0.70\columnwidth}{!}{\includegraphics{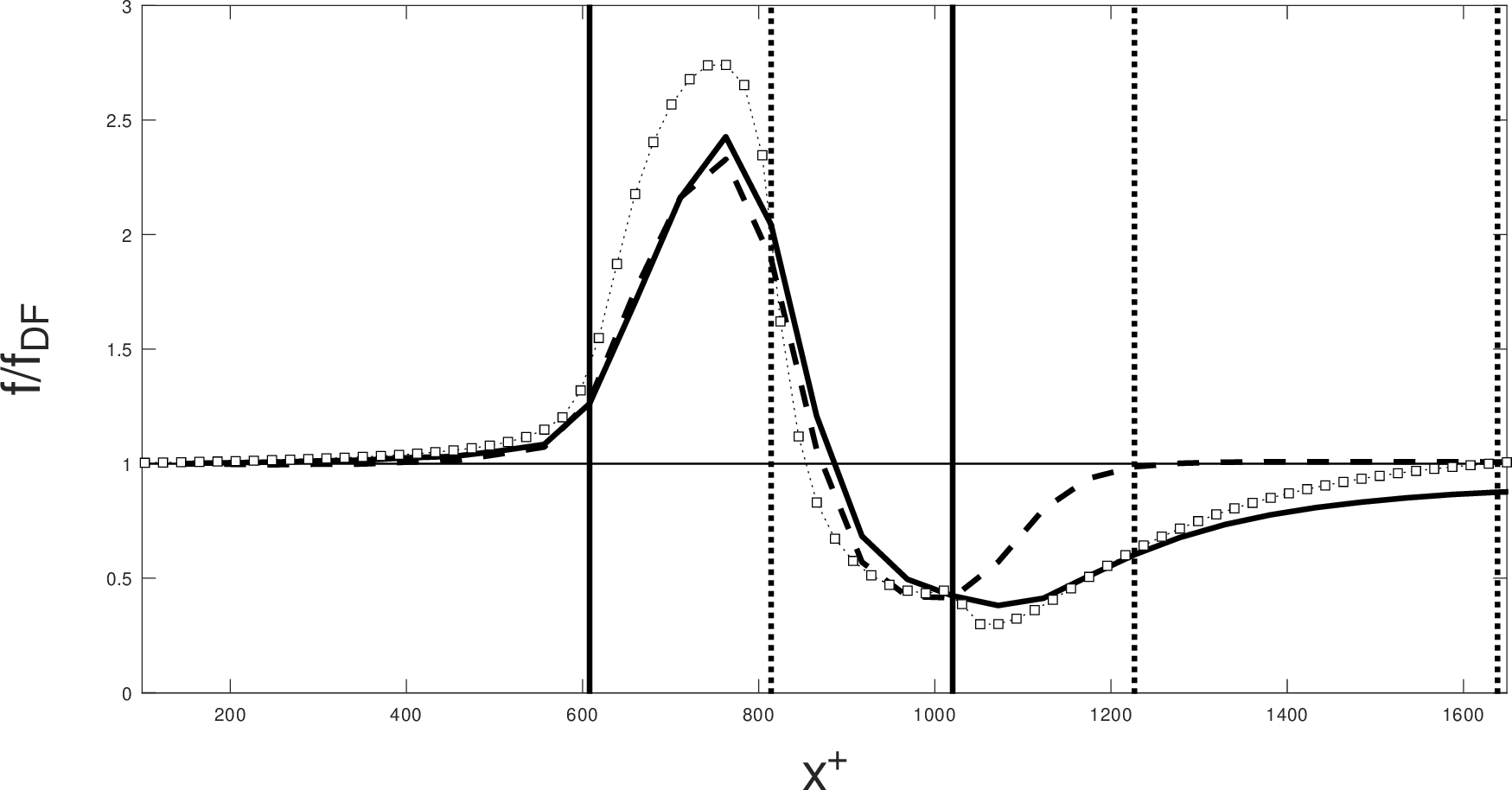}}
  \par}
\caption{\small{Normalized skin-friction coefficient $f/f_{DF}$ for a
    channel flow with $Re_{\tau}=300$, perturbed by wall suction
    followed by blowing in adjacent regions, each with a width of $W^+
    = 206$, for a total perturbation width of $W^+ =
    412$. \brokenline, \normalsize $\kappa-\epsilon$ model; \boxs
    \normalsize, DNS; solid line, LSTM.}}
\label{figure4} 
\end{figure}

Since the turbulent flows analyzed in this study are symmetric in the
z-direction, they are treated as two-dimensional statistical turbulent
flows. As a result, the only significant shear stress is $\langle uv
\rangle$, which describes the transfer of momentum in the x-direction
from the channel's central region (higher momentum) toward the wall
(lower momentum). The normal stresses are considered not relevant in
this context and are set to zero to simplify the LSTM
model. Consequently, the LSTM model is designed to predict and
propagate only the $\langle uv \rangle$ shear stress. In contrast, the
$\kappa-\epsilon$ model accounts for all Reynolds stresses, with none
assumed to be zero.

\begin{figure}[tbp] 
{\par\centering
  \resizebox*{0.70\columnwidth}{!}{\includegraphics{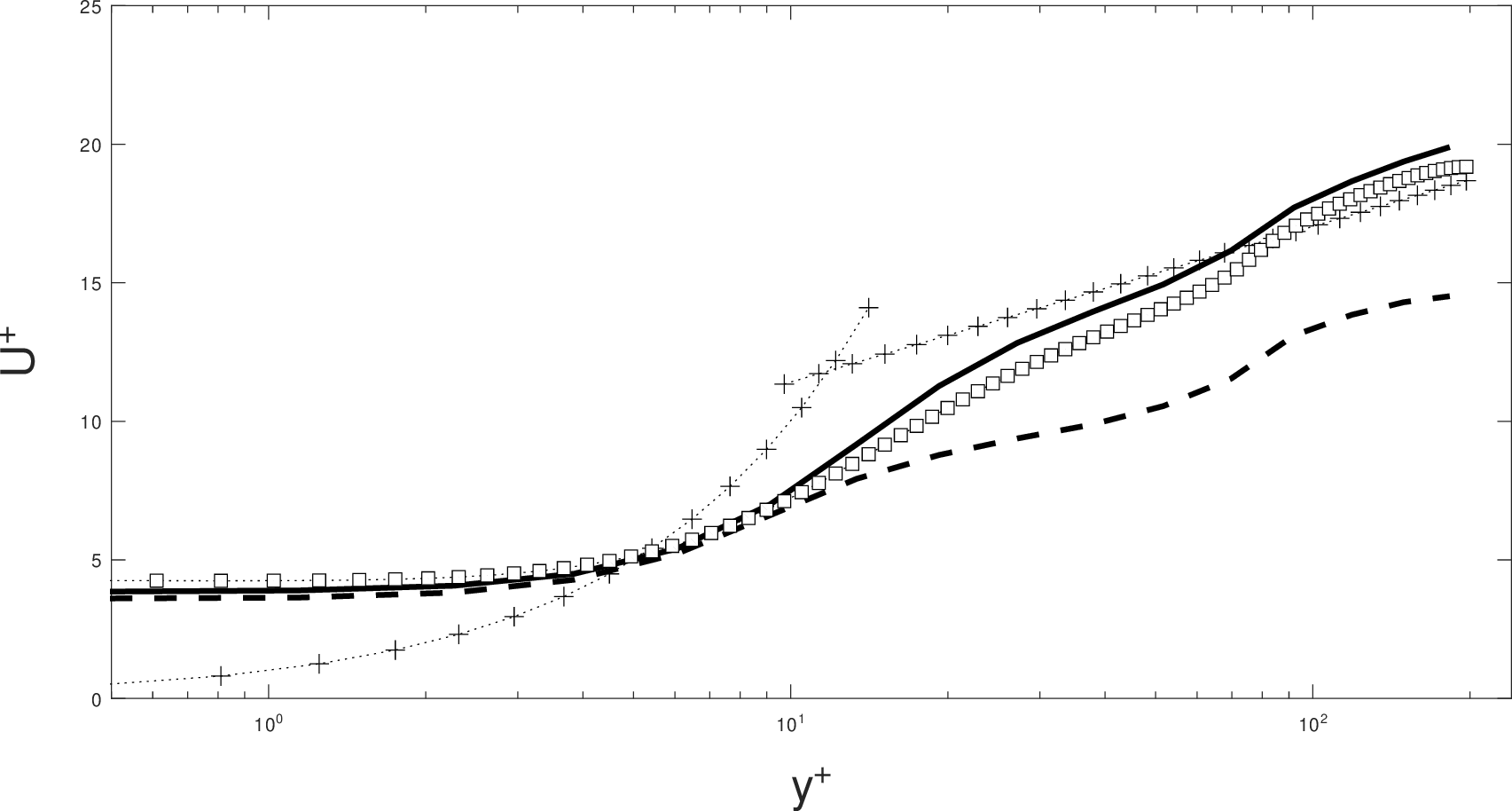}}
  \par}
\caption{\small{Dimensionless longitudinal velocity $U^+$ at the
    center of the perturbation region for a channel flow with
    $Re_{\tau}=200$, subjected to a frictionless wall and an adverse
    pressure gradient step in a narrow region of width
    $W^+=206$. \boxs \normalsize, DNS; \brokenline \normalsize,
    $\kappa-\epsilon$ model; solid line, LSTM; \pluss \normalsize, Law
    of the Wall, $U^+ = ({1}/{0.41})\ln(y^+) + 5.5$.}}
\label{figure5} 
\end{figure}

\begin{figure}[tbp] 
{\par\centering
  \resizebox*{0.70\columnwidth}{!}{\includegraphics{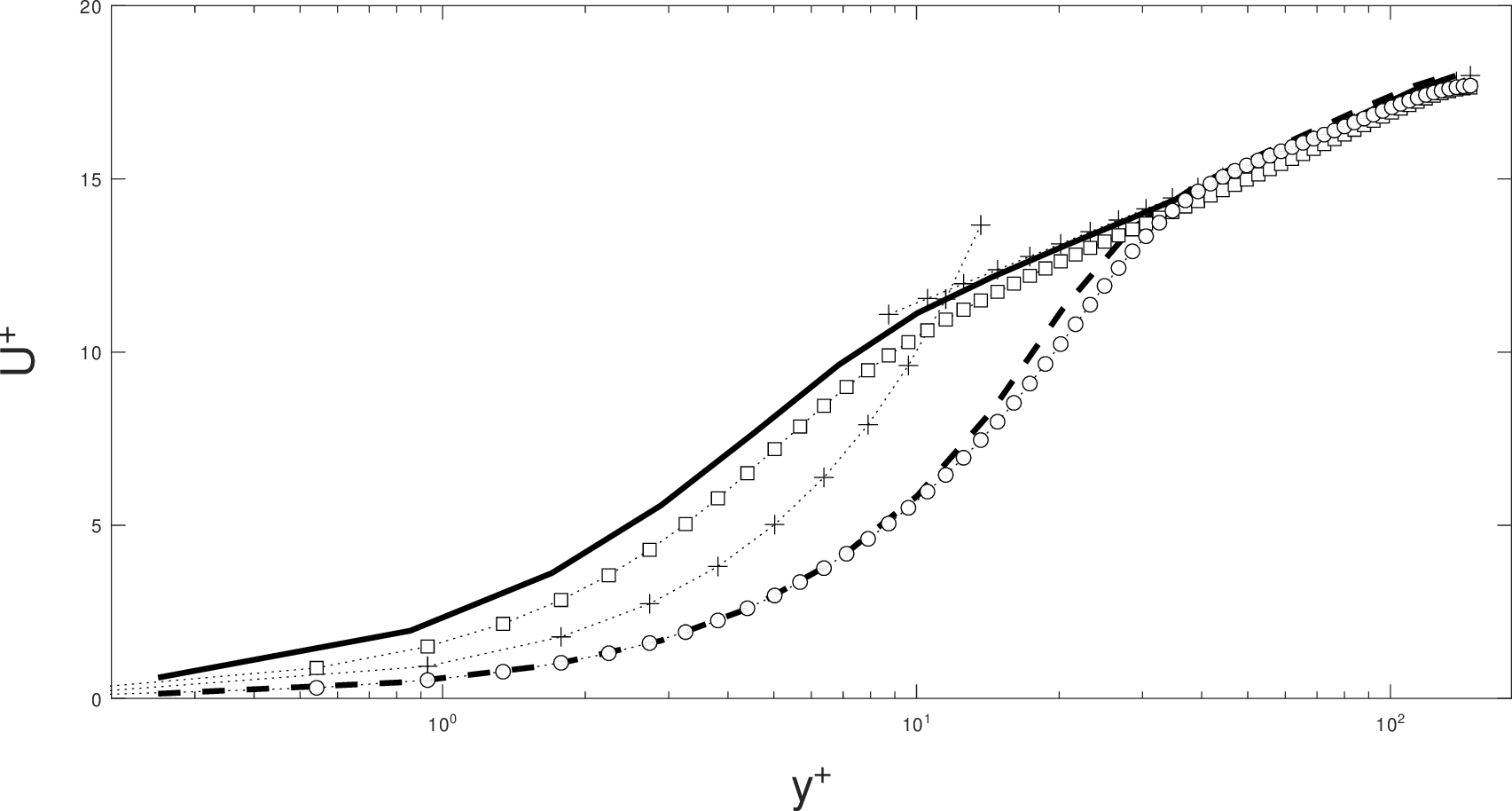}}
  \par}
\caption{\small{Dimensionless longitudinal velocity $U^+$ for a
    channel flow with $Re_{\tau}=150$, perturbed by wall suction
    followed by blowing, in a narrow region of total width
    $W^+=412$. Middle of the perturbation: \boxs \normalsize, DNS;
    solid line, LSTM. At a distance of $W^+=206$ from the end of the
    perturbation: \circs \normalsize, DNS; \brokenline \normalsize,
    LSTM; \pluss \normalsize, Law of the Wall, $U^+ =
    ({1}/{0.41})\ln(y^+) + 5.5$.}}
\label{figure6} 
\end{figure}

\begin{figure}[tbp] 
{\par\centering
  \resizebox*{0.70\columnwidth}{!}{\includegraphics{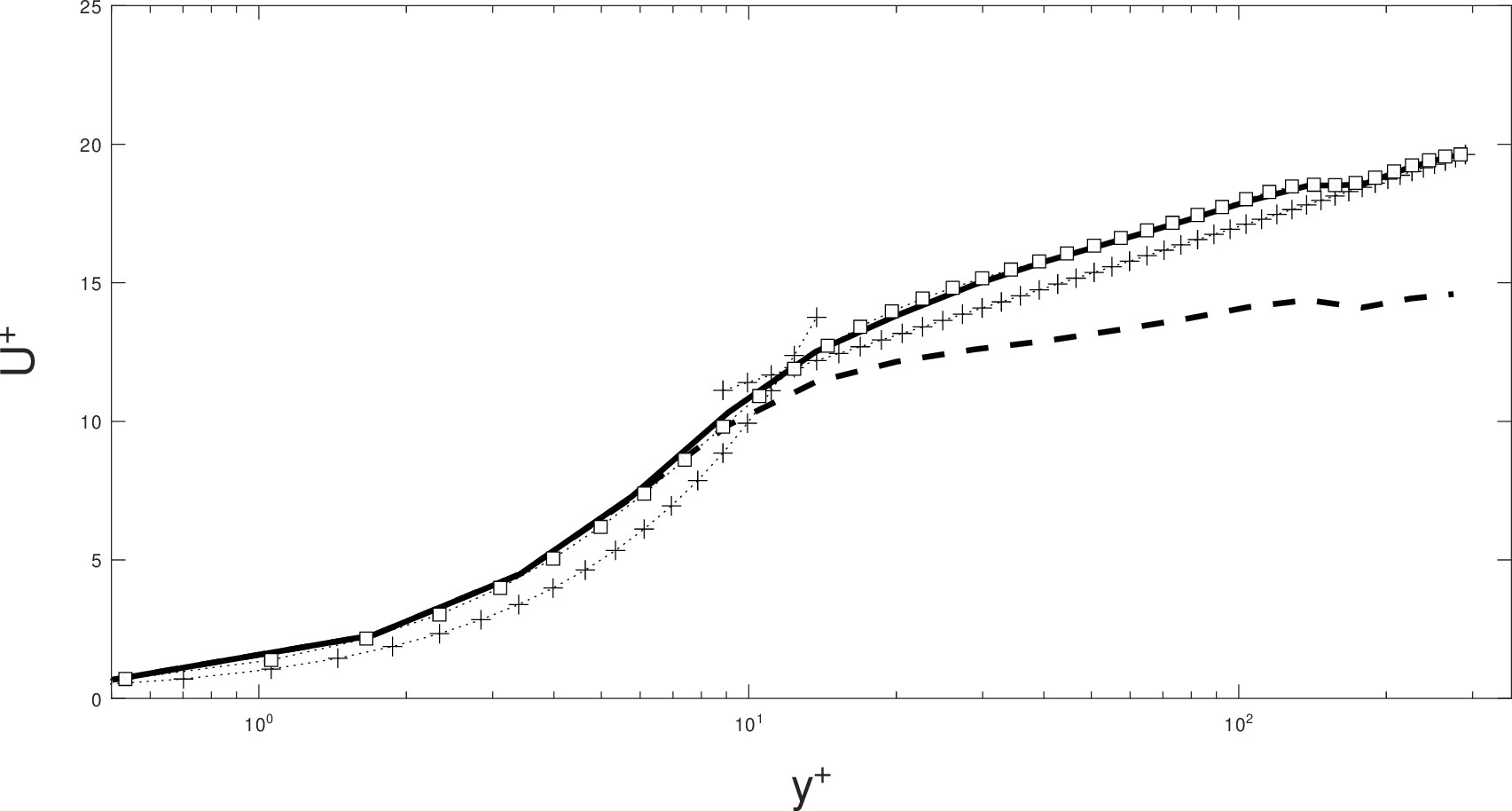}}
  \par}
\caption{\small{Dimensionless longitudinal velocity $U^+$ at a
    $W^+=206$ distance from the end of the perturbation region for a
    channel flow with $Re_{\tau} \simeq 300$, perturbed by a
    frictionless wall and a favorable pressure gradient in a narrow
    region of width $W^+=206$. \boxs \normalsize, DNS; \brokenline
    \normalsize, $\kappa-\epsilon$ model; solid line, LSTM; \pluss
    \normalsize, Law of the Wall, $U^+ = ({1}/{0.41})\ln(y^+) +
    5.5$.}}
\label{figure7} 
\end{figure}

This study focuses on moderately low Reynolds number turbulent flows,
with the maximum friction Reynolds number $Re_{\tau}$ being 600, which
corresponds to $Re_{\delta} \simeq 14,000$. The results serve as a
proof of concept, demonstrating the potential for LSTMs to replace
traditional RANS turbulence models. However, further research is
necessary to extend this approach to higher Reynolds numbers and more
complex flow conditions. As mentioned earlier, creating a universal
LSTM model that works across a wide range of Reynolds numbers and flow
characteristics would likely require a large number of parameters,
which could limit its practicality for CFD applications. An
alternative approach may involve using specialized neural networks
tailored to specific types of turbulent flows. For example, an LSTM
model trained exclusively on a DNS database of fully developed
turbulent flows over a range of Reynolds numbers could be more
efficient for such cases. There would be no need for a more complex
LSTM with additional parameters trained on unrelated flow types.

The LSTM predictions are compared with results from the
$\kappa-\epsilon$ model and DNS data. The statistics presented include
longitudinal velocity ($U$) and shear Reynolds stress (turbulent,
viscous, and total) for developed flows, as well as wall skin
friction, longitudinal mean velocity, and $\langle uv \rangle$ for
perturbed flows, all in dimensionless form.

\subsection{Developed Flow}

As mentioned earlier, all results presented in this study, using the
LSTM neural network, correspond to the 4XBL13BL23BL33DL
architecture. Figures \ref{figure1} and \ref{figure2} display the
mean longitudinal velocity, $U^+$, for various Reynolds numbers and
the Reynolds shear stress $\langle uv \rangle^+$ for developed flow at
$Re_{\tau}=200$ and $Re_{\tau}=600$, respectively (both results in
dimensionless form using the friction velocity, $u_{\tau}$). These
figures highlight the LSTM's capability to accurately predict the
Reynolds shear stress in developed flows, resulting in a correct mean
velocity distribution, with good agreement with the 'Law of the Wall'.

The input features for the 4XBL13BL23BL33DL architecture, as listed in
Table \ref{table1}, include $Re_{\tau}$, $S_{11}^+$, $S_{12}^+$, and
$Y$. Here, the plus symbol indicates that these features are
nondimensionalized using wall parameters $u_{\tau}$ and $\delta$,
except for the wall distance $Y$, which is nondimensionalized by
$\delta$ (instead of the usual $y^+=y u_{\tau}/ \nu$). The inclusion
of $Re_{\tau}$ as a feature was motivated by the improvement it
provided in predicting $\langle uv \rangle^+$ for developed flows. As
$Re_{\tau}$ increases, the slope of $\langle uv \rangle^+$ in the
near-wall region becomes steeper (Fig. \ref{figure2}), making the
inclusion of $Re_{\tau}$ essential for accurate predictions under
developed flow conditions.

\subsection{Skin-Friction for Perturbed Flow}

Figures \ref{figure3} and \ref{figure4} present the
normalized skin-friction coefficient, $f/f_{DF}$ (normalized by the
skin-friction coefficient for developed flow, $f_{DF}$), for the RANS
simulations based on propagated LSTM predictions, compared with the
$\kappa-\epsilon$ model and DNS data. It is important to note that all
RANS simulations (both for LSTM propagation and the $\kappa-\epsilon$
model) were performed using the same numerical grid, with identical
distances from the wall to the center of the first cell in the grid.

From an applied engineering perspective, the primary objective of a
RANS simulation is to accurately compute the mean flow, with
particular emphasis on predicting wall skin friction. The shear
Reynolds stress, $\langle uv \rangle^+$, is crucial in determining the
near-wall mean flow distribution, as it influences the wall-normal
gradient of the mean velocity. Accurate prediction of $\langle uv
\rangle^+$ near the wall is therefore essential. In aerodynamic
analysis, for example, wall skin friction is a key parameter due to
its direct relationship with fuel consumption and other critical
factors in vehicle performance evaluation.

Figure \ref{figure3} shows the abrupt change in $f/f_{DF}$ for a
channel flow originally in hydrodynamically developed conditions at
$Re_{\tau}=200$, perturbed in a narrow region with width $W^+=206$,
where a frictionless wall and an adverse pressure gradient step causes
the skin friction to drop to zero. This type of perturbation abruptly
alters the turbulent momentum fluxes from the channel center toward
the wall, strongly affecting the wall skin friction.

Figure \ref{figure4} presents $f/f_{DF}$ for a developed channel
flow at $Re_{\tau}=300$, perturbed by suction from the wall in a
narrow region of width $W^+=206$, followed by blowing in a similar
region. The suction first increases the skin friction as fluid is
drawn toward the wall, enhancing the wall-normal mean velocity near
the surface. In contrast, the subsequent blowing lifts the fluid away
from the wall, causing a sharp decrease in the skin-friction
coefficient.

In both figures, the RANS simulations follow the general trend
observed in the DNS data, despite the coarse numerical grid used. The
performance of the LSTM model surpasses that of the $\kappa-\epsilon$
model, generally providing better alignment with the DNS data outside
the perturbation region. To further improve the LSTM's skin-friction
predictions, a denser numerical grid would likely be more effective
than merely increasing the number of parameters in the neural network.

\subsection{Mean Velocity for Perturbed Flow}

Figures \ref{figure5} to \ref{figure7} illustrate the
distribution of the longitudinal mean velocity, $U^+$, for various
cases of perturbed channel flow. In Fig. \ref{figure5}, $U^+$
is shown at the center of the perturbation region for a flow at
$Re_{\tau}=200$, subjected to a frictionless wall and an adverse
pressure gradient step. The frictionless wall significantly alters the
pure adverse pressure gradient case used for LSTM model training by
suspending the turbulent transfer of longitudinal momentum toward the
wall. As a result, the mean velocity is modified across the entire
channel, and $U^+$ no longer follows the 'Law of the Wall' throughout
the domain.

In Fig. \ref{figure6}, the mean velocity predicted by LSTM
propagation is compared to DNS data for a channel flow at
$Re_{\tau}=150$, perturbed by wall suction followed by blowing within
a narrow region of width $W^+=412$. This perturbation creates an
abrupt increase in the velocity gradient near the wall during suction,
followed by a sharp decrease in the gradient during blowing (the
wall-normal velocity gradient peaks in the suction region and reaches
a minimum in the adjacent blowing region). In this case, the mean flow
is substantially modified for $y^+ < 50$, and $U^+$ approaches the
'Law of the Wall' from the midpoint to the channel center. It is
evident that $U^+$ exceeds the 'Law of the Wall' mean velocity at the
end of the suction region but falls below it at a distance $W^+$ from
the blowing end. Essentially, suction draws fluid with higher
longitudinal momentum toward the wall, increasing the mean velocity in
the near-wall region, whereas blowing blocks the mean flow near the
wall, reducing the longitudinal mean velocity.

Figure \ref{figure7} presents $U^+$ at a distance $W^+$ from
the end of the perturbation region for a flow perturbed by a
frictionless wall and a favorable pressure gradient step, at
$Re_{\tau}=300$. In this case, the mean velocity slightly exceeds the
developed condition (surpassing the 'Law of the Wall') throughout most
of the channel, except near the center.

All these figures highlight the strong performance of the
LSTM-predicted mean flow, with results closely aligning with DNS data
and generally outperforming the $\kappa-\epsilon$ model.

\subsection{Shear Reynolds Stress}

\begin{figure}[tbp]
  {\par\centering
  \resizebox*{0.70\columnwidth}{!}{\includegraphics{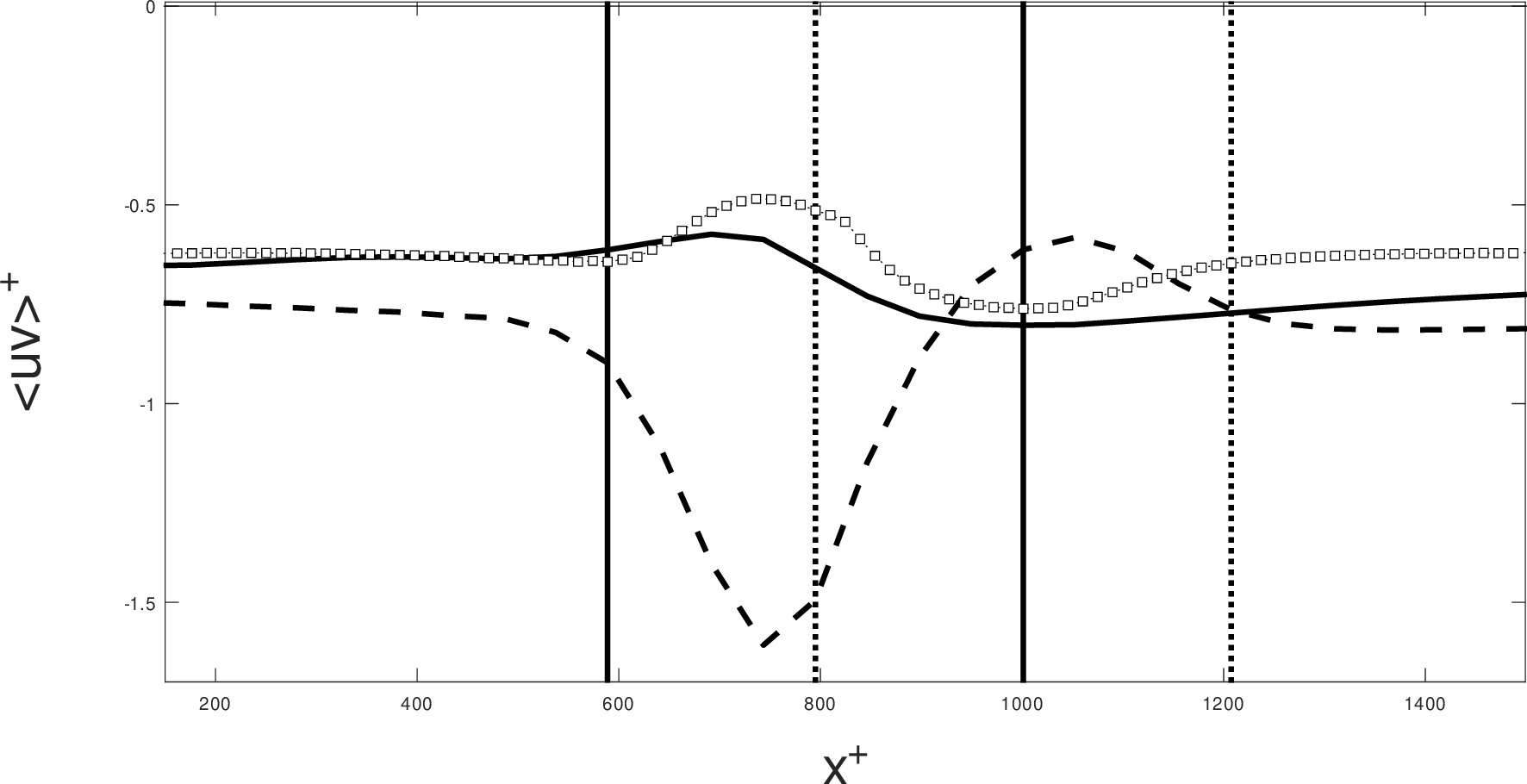}}
  \par}
  \caption{\small{$\langle uv \rangle^+$ along $X^+$ at $Y^+\simeq 17$
      from the wall, for a channel flow with $Re_{\tau} \simeq 150$,
      perturbed with blowing followed by suction in a narrow region of
      width $W^+=412$. \brokenline \normalsize $\kappa-\epsilon$;
      \boxs \normalsize, DNS; solid line, LSTM.}}
  \label{figure8}
\end{figure}

\begin{figure}[tbp]
  {\par\centering
  \resizebox*{0.70\columnwidth}{!}{\includegraphics{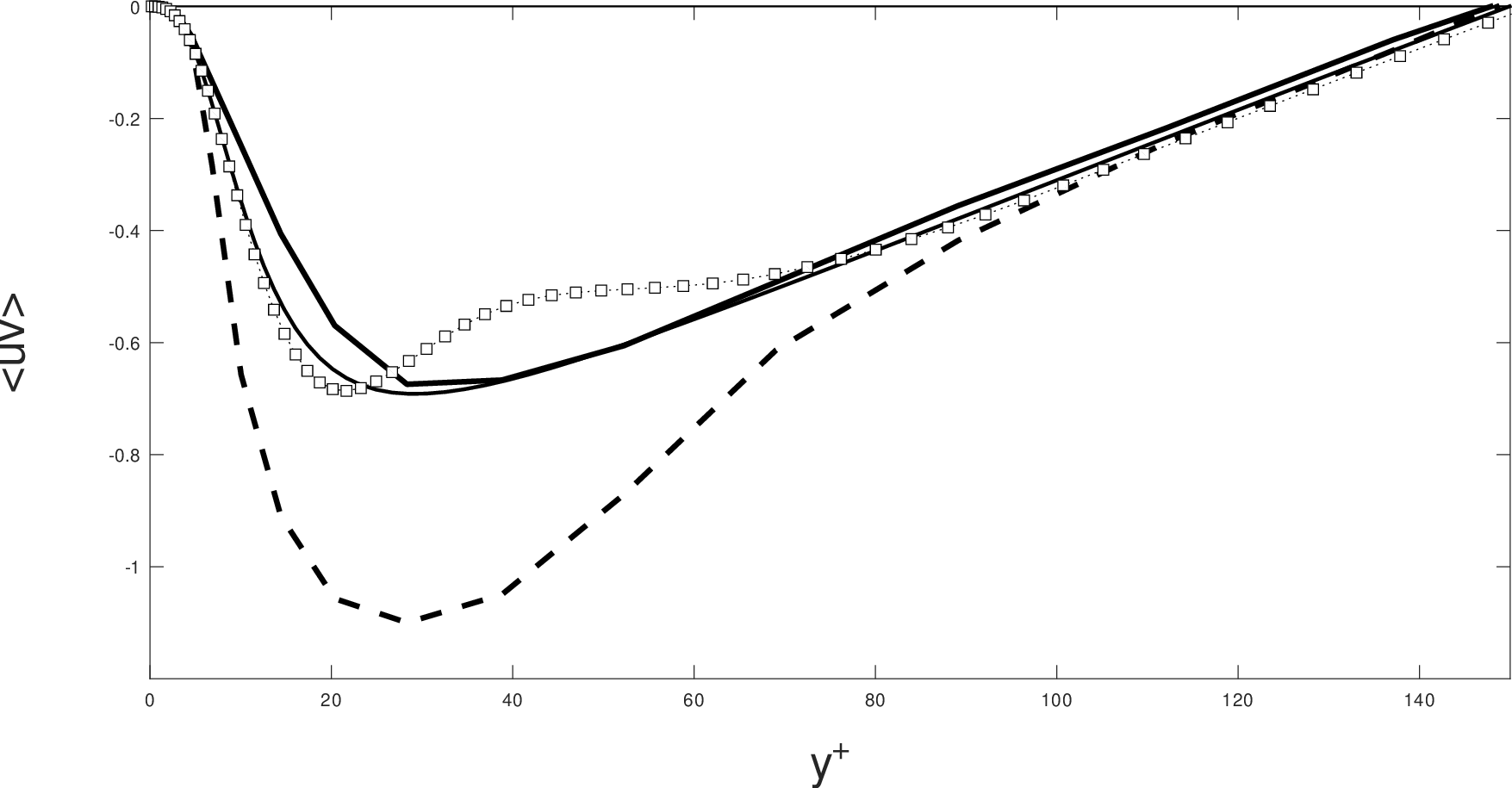}}
  \par}
  \caption{\small{Wall-normal distribution of $\langle uv \rangle^+$
      at a distance $W^+$ from the perturbation end for a flow with
      $Re_{\tau} \simeq 150$ perturbed by wall suction followed by
      blowing in a region with a total width of
      $W^+=412$. \brokenline, \normalsize $\kappa-\epsilon$ model;
      \boxs \normalsize, DNS; thick solid line, LSTM; thin solid line,
      DNS in developed condition.}}
  \label{figure9}
\end{figure}

\begin{figure}[tbp]
  {\par\centering
  \resizebox*{0.70\columnwidth}{!}{\includegraphics{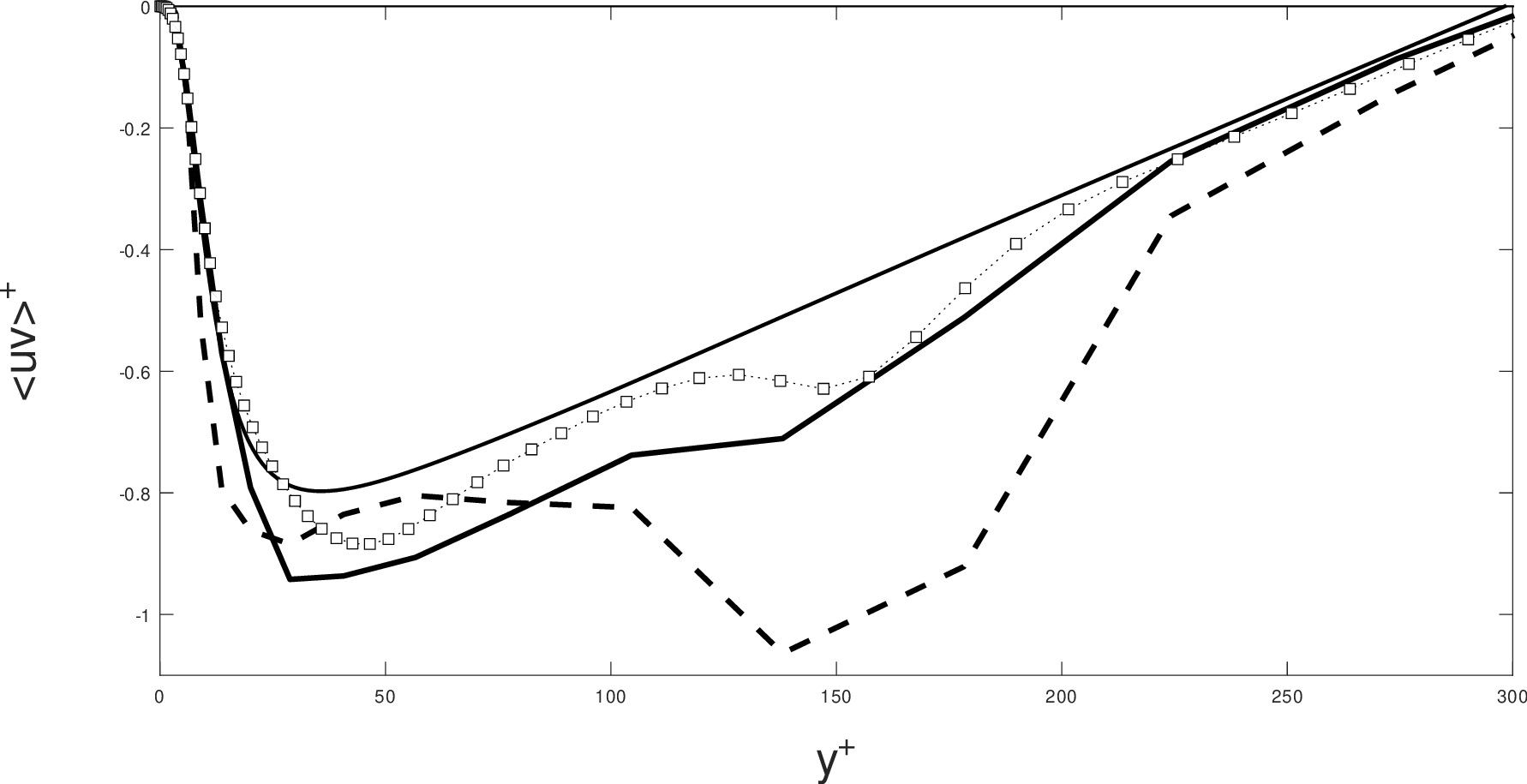}}
  \par}
  \caption{\small{Wall-normal distribution of $\langle uv \rangle^+$
      at $W+=206$ from the perturbation end for a flow with $Re_{\tau}
      \simeq 300$ perturbed by a frictionless wall and an adverse 
      pressure gradient step. \brokenline,
      \normalsize $\kappa-\epsilon$ model; \boxs \normalsize, DNS;
      thick solid line, LSTM; thin solid line, DNS in developed condition.}}
  \label{figure10}
\end{figure}

\begin{figure}[tbp]
  {\par\centering
    \resizebox*{0.70\columnwidth}{!}{\includegraphics{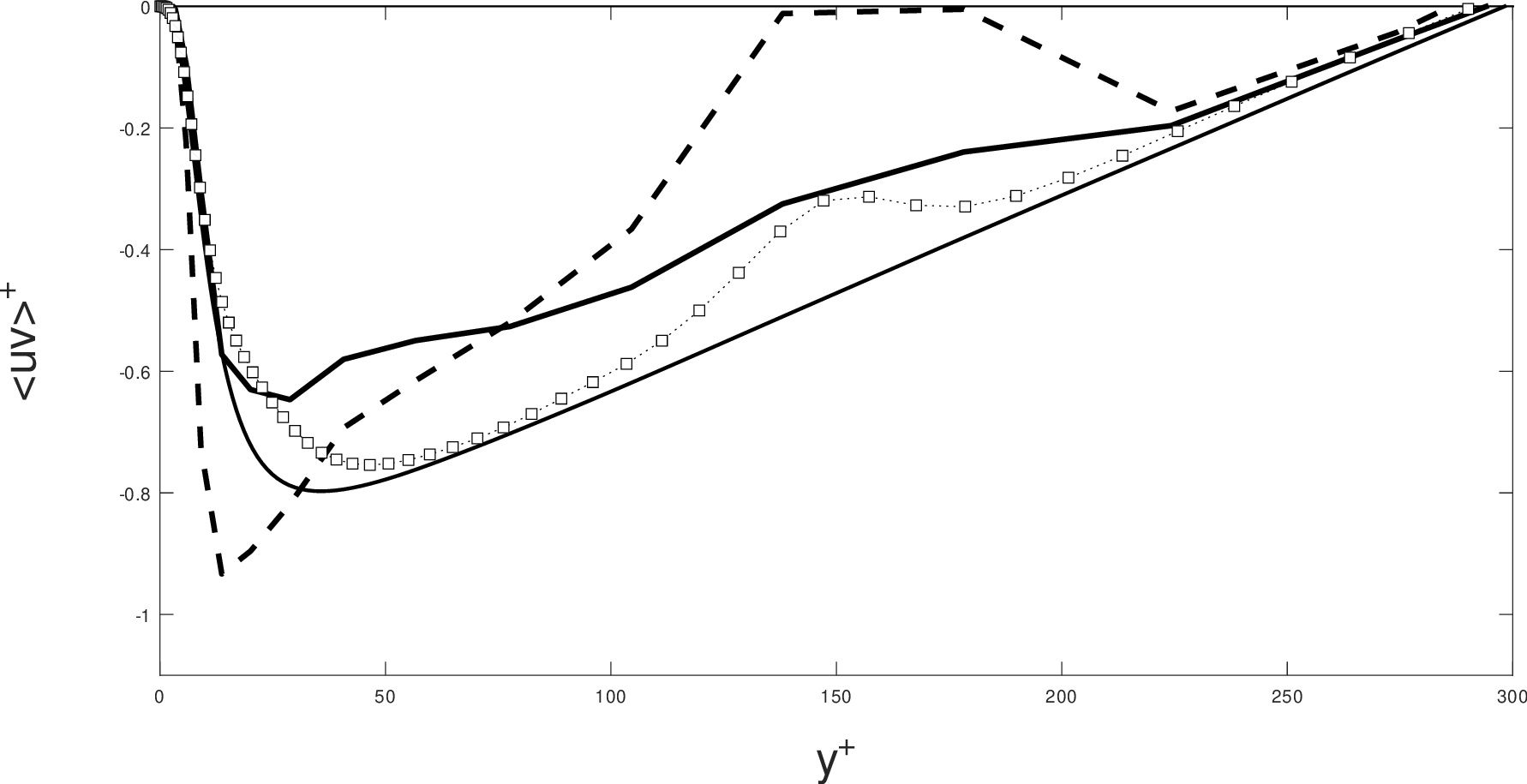}}
    \par}
  \caption{\small{Wall-normal distribution of $\langle uv \rangle^+$
      at $W+=206$ from the perturbation end for a flow with $Re_{\tau}
      \simeq 300$ perturbed by a frictionless wall and a favorable
      pressure gradient step. \brokenline, \normalsize
      $\kappa-\epsilon$ model; \boxs \normalsize, DNS; thick solid
      line, LSTM; thin solid line, DNS in developed condition.}}
  \label{figure11}
\end{figure}

As previously mentioned, the RANS simulations based on the LSTM
predictions in this study use only the shear Reynolds stress, $\langle
uv \rangle^+$, while the other Reynolds stresses are set to
zero. Figures \ref{figure8} through \ref{figure11}
compare the LSTM’s predictions of $\langle uv \rangle^+$ with DNS data
and results from the $\kappa-\epsilon$ model.

The aim of this study is not to critique the $\kappa-\epsilon$ model,
which has long been a benchmark for researchers working with RANS
simulations. Despite its simplicity, the $\kappa-\epsilon$ model has
provided reasonable predictions for a wide range of turbulent flows
over the last 50 years. However, it is well-known that the model tends
to overestimate turbulence kinetic energy and, consequently,
overpredicts $\langle uv \rangle^+$. This excessive dissipation leads
to an overestimation of the Reynolds shear stress compared to DNS
data.

In contrast, while the LSTM predictions do not always align perfectly
with DNS data, they consistently exhibit errors due to
underestimation. For cases where the absolute value of $\langle uv
\rangle^+$ exceeds the developed condition, the LSTM model follows
this trend but predicts lower absolute values than the DNS
results. Similarly, for cases where $\langle uv \rangle^+$ is lower
than its developed condition, the LSTM predictions again follow this
behavior but with reduced absolute values.

In Fig. \ref{figure8}, the $\langle uv \rangle^+$ distribution
along the longitudinal direction at $y^+ \simeq 17$ (where turbulence
production is significant) shows that the LSTM model agrees better
with DNS data compared to the $\kappa-\epsilon$ model. However, the
agreement remains relative, with the LSTM prediction slightly
underestimating the DNS data.

A similar trend is observed in Figs. \ref{figure9} to
\ref{figure11}, which depict the wall-normal distributions of
$\langle uv \rangle^+$ for different cases. In
Fig. \ref{figure9}, the DNS data almost matches the developed
flow value, and the LSTM model shows minimal deviation from the
developed condition, while the $\kappa-\epsilon$ model predicts an
excessive value. This conservative behavior of the LSTM model can be
viewed as a strength, providing reliable and stable results.

Figures \ref{figure10} and \ref{figure11} illustrate
cases with perturbations involving a frictionless wall and either an
adverse or favorable pressure gradient. These perturbations have
opposite effects on the turbulent flow, as reflected in the DNS
data. The LSTM predictions follow the same trends but in a more
conservative manner, while the $\kappa-\epsilon$ model predicts
extreme values of $\langle uv \rangle^+$ that significantly exceed DNS
results.

Overall, these figures demonstrate the consistent behavior of the LSTM
model, which remains conservative compared to the more extreme
overpredictions of the $\kappa-\epsilon$ model.

\section{Conclusions}

This second part of the study (the preprint of the first part is
titled 'Modeling Turbulent Flows with LSTM Neural Network',
arXiv:2307.13784v1 [physics.flu-dyn], 25 July 2023) delves deeper into
the application of the Long Short-Term Memory (LSTM) artificial
recurrent neural network (RNN) as an alternative to traditional
Reynolds-Averaged Navier-Stokes (RANS) models. The LSTM model was
employed to predict shear Reynolds stress in both developed and
developing turbulent channel flows, and these predictions were then
propagated in RANS simulations to obtain mean flow solutions. A
comparative analysis was performed between LSTM-based results from
computational fluid dynamics (CFD) simulations, the outcomes from the
$\kappa-\epsilon$ model, and data from direct numerical simulations
(DNS). These analyses highlighted the promising performance of the
LSTM approach.

The main conclusions of this study, which employed an LSTM
architecture with approximately 700 parameters tailored to the
characteristics of statistical 2D turbulent channel flows and used as
a RANS turbulence model, are as follows.

The LSTM's propagated predictions were robust, as the turbulent stress
values predicted by the model were integrated into computational fluid
dynamics (CFD) codes in a stable and reliable manner, at least for the
moderately low Reynolds numbers considered in this
study. Additionally, the LSTM demonstrated efficacy in non-developed
flows, with new RANS simulation cases revealing the model’s
effectiveness in capturing turbulence dynamics in perturbed turbulent
flows. Despite using a coarse numerical grid in the RANS propagation,
the LSTM predictions maintained a reasonable level of accuracy across
all cases. Moreover, the LSTM exhibited a conservative behavior, which
proved advantageous by providing reliable and stable predictions in
comparison to traditional turbulence models.

While this study primarily serves as a proof-of-concept for employing
LSTM neural networks in RANS turbulence modeling, this second phase
provides strong evidence of its potential as an alternative to
conventional models.

\vskip1cm

\noindent\textbf{Acknowledgments}

\vskip0.3cm

The author gratefully acknowledge the developers of TensorFlow and
Keras for providing these open-source libraries, which were invaluable
for the implementation of the machine learning models in this
work. The availability of these resources has greatly facilitated the
research presented in this preprint.

\bibliographystyle{unsrtnat}
\bibliography{LSTMforRANS}

\end{document}